\documentclass{article}
\usepackage{graphicx,  bm, amsmath, amsfonts, amssymb, amsthm, geometry} 
\usepackage[
authoryear, 
]{natbib}
\usepackage{csquotes}

\usepackage{color}
\definecolor{darkgreen}{rgb}{0, .5, 0}
\definecolor{darkred}{rgb}{.5, 0, 0}

\newcommand{\interior}[1]{%
  {\kern0pt#1}^{\mathrm{o}}%
}

\newcommand{\Levy}{L\'{e}vy }

\newcommand{\cadlag}{c\`adl\`ag}

 \newcommand{\T}{\mathcal{T}}

\newcommand{\BSS}{BSS }

\newcommand{\R}{\ensuremath{\mathbb{R}}}
\newcommand{\N}{\ensuremath{\mathbb{N}}}

\newcommand{\EE}{\ensuremath{\mathbb{E}}}

\newcommand{\Var}{\ensuremath{\mathrm{Var}}}

\newcommand{\RV}{\ensuremath{\mathrm{RV}}}
\newcommand{\SRV}{\ensuremath{\mathrm{SRV}}}

\newcommand{\Prob}{\ensuremath{\Bbb{P}}}

\newcommand{\leb}{\ensuremath{Leb}}

\newcommand{\indicator}{\ensuremath{\mathbb{I}}}
\def\e{{\text{e}}}

\newcommand{\ucp}{\ensuremath{\stackrel{\text{u.c.p.}}{\rightarrow}}}

\newcommand{\Yambit}{Y(\mathbf{x},t)}

\newcommand{\Aambit}{A(\mathbf{x},t)}

\newtheorem{assumption}{Assumption}[section]
\newtheorem{theorem}{Theorem}[section]
\newtheorem{corollary}{Corollary}[section]
\newtheorem{remark}{Remark}[section]
\newtheorem{example}{Example}[section]

\title{Research frontiers in ambit stochastics:\\ In memory of Ole E.~Barndorff-Nielsen}
\author{Fred Espen Benth\footnote{Department of Mathematics, University of Oslo, P.O. Box 1053 Blindern, 0316 Oslo, Norway, fredb@math.uio.no} \and Almut E.~D.~Veraart\footnote{Department of Mathematics, Imperial College London, 180 Queen's Gate, London, SW7 2AZ, UK, a.veraart@imperial.ac.uk} }
\date{}

\begin{document}

\maketitle

\begin{abstract}
This article surveys key aspects  of  ambit stochastics and remembers Ole E.~Barndorff-Nielsen's important contributions to the foundation and advancement of this new research field over the last two decades. It also highlights some of the emerging trends in ambit stochastics. 
\end{abstract}



\section{Introduction}
\label{intro:sec:1}
This article remembers and celebrates
Ole E.~Barndorff-Nielsen's contributions to ambit stochastics and points towards recent developments in this field. 

Throughout his research, Ole was fascinated by developing stochastic models for turbulence that align with the \emph{Statistical Theory of Turbulence} as described by Kolmogorov and Obukhov, see \cite{Kol41b, Obu41} and also \cite{Frisch95}.
This endeavour led to the introduction of a new field in probability, termed ambit stochastics.

Ambit fields and processes are powerful tools for modelling random phenomena in space-time, see the relevant survey papers  
 \cite{BNBV2014,
Podolskij2014, BHSS2016,
Schmiegel2018}. Also, the recent monograph
\cite{BNBV2018_ambitbook} gives a comprehensive summary of some of the key results in the area of ambit stochastics that were developed until 2018. 

Let us briefly review the definition of an ambit field. 
Let $Y=(\Yambit)$ denote a random field taking values in $\R$. 
Here, $\mathbf{ x}\in \R^d$ ($d\in \N$) and $t\in \R$ are coordinates in space and time, respectively. For instance, we can think of $Y$ describing the wind speed at time $t$ in location $\mathbf x$.
Since general random fields are often difficult to deal with, we typically impose some structural or model assumptions to achieve mathematical tractability. At the same time, we want to ensure that the model class we consider is very flexible and can describe empirical data. 
Ambit fields provide a flexible, yet analytically tractable class of random field models, which go far beyond the widely used Gaussian random fields.

The key component of an ambit field takes the form 
\begin{equation}\label{eq:Cha0_ambitfield}
\Yambit=\int_{\Aambit }g(\mathbf{ x},t;\boldsymbol{\xi},s)\sigma(\boldsymbol{ \xi}, s)L_{\T}(d\boldsymbol{\xi}, ds).
\end{equation}

Here, $\Aambit\subseteq \R^d\times \R$ is referred to as the ambit set which can vary with the space-time coordinates  $(\mathbf x, t)$.
Moreover, the integrand consists of a product of a deterministic kernel function  $g$ and a stochastic volatility/intermittency field  $\sigma$. Finally,  $L$ is a L\'{e}vy basis, that is, a particular random noise term which generalises (Gaussian) white noise and \Levy processes to space-time settings.
For some applications, it might be useful to consider a \Levy basis which is subordinated by a metatime $\T$, see \cite{BNPedersen2012}, in which case we write $L_{\T}$. 

\subsection{Stochastic volatility/intermittency}
Ole was particularly interested in the concept of stochastic volatility (a term widely used in financial econometrics) and/or intermittency (a term more dominant in the turbulence literature). Volatility/intermittency are relative concepts and need to be understood as additional fluctuations besides those one would otherwise anticipate. Here the stochastic volatility/intermittency field/process $\sigma$ 
introduces such additional variability via spatial scaling of the \Levy noise $L$. 
What is more, Ole always stressed the importance of modelling stochastic volatility/intermittency via stochastic time change or, more generally, the concept of metatimes.

Ole contributed widely to the literature on stochastic time change for stochastic processes, see, amongst others, \cite{Barndorff-Nielsen_Pedersen_Sato_2001, BNS2001, BNPA2008} and wrote the influential book on \emph{Change of Time and Change of Measure}, see  \cite{BNShiryaev2010}. 
In order to also cover spatial and spatio-temporal settings, Ole introduced the concept of metatimes in the paper \cite{BNPedersen2012}.

Early work in the area of ambit stochastics focusses predominantly on modelling questions. For instance, together with J\"urgen Schmiegel, Ole had a long-standing collaboration on working on stochastic models for turbulence, with the first paper mentioning the ideas of ambit fields in this context being  \cite{BNSch07a}.





\subsection{The role of the ambit set and the kernel function}
A key feature in the definition of an ambit field is the 
 \emph{ambit set} $\Aambit$. 
 The term \emph{ambit} has been derived from the  Latin word \emph{ambitus} which means \emph{border, boundary or sphere of influence}. Here, at each point in space-time $(\mathbf x,t)$, the ambit set describes the relevant region over which random noise (described via the \Levy basis $L$) needs to be taken into account to model $\Yambit$.   
 Intuitive examples of ambit sets include the light or sound cone. 
 
 The reader might wonder, why ambit fields were defined as in \eqref{eq:Cha0_ambitfield}
 rather than using the simpler expression
\begin{equation*}
\Yambit=\int G(\mathbf x,t;\boldsymbol \xi,s)L_{\T}(d\boldsymbol\xi, ds),
\end{equation*}
where $G(\mathbf x,t;\boldsymbol \xi,s)=\indicator_{\Aambit }(\boldsymbol \xi,s)g(\mathbf x,t;\boldsymbol \xi,s)\sigma(\boldsymbol \xi, s)$. 
The reason for including the ambit sets explicitly is that they give the modeller a means to directly include any boundaries for the sphere of influence of the random fluctuations in space-time.
Also, by separating the deterministic kernel function $g$ from a stochastic volatility component $\sigma$, we will later be able to derive tailor-made inference approaches for various applications of interest. 
That said, it will always be important to investigate the identifiability of the particular models used.

\subsection{Outline}
The remainder of this article is structured as follows. Section \ref{sec:stochvolest} surveys some of the key results on statistical inference for stochastic volatility in ambit fields and related processes. 
Section \ref{sec:SPDElinks} discusses the important relation between ambit fields and stochastic partial differential equations (SPDEs) and surveys some of the recent advances in volatility estimation for infinite-dimensional settings, directly extending results discussed in Section \ref{sec:stochvolest}. 
Next, we showcase some of the key application areas of ambit fields in 
Section \ref{sec:practice}, and 
Section \ref{sec:con} concludes and gives an outlook on some of the latest advances in ambit stochastics.

\section{Statistical inference for stochastic volatility and intermittency}\label{sec:stochvolest}
We have discussed the importance of including stochastic volatility in the modelling and will now study the question of how  volatility can be  estimated. 
Crucially, the answer to this question will depend on the particular structure of the process and we will pick three stylised cases to explain the general theory; in all these cases we will focus on ambit processes driven by a Brownian motion and will give references to results in the \Levy case below.  

We will begin our study with a purely  temporal setting focussing on Brownian semistationary processes. Here we need to distinguish two cases: the semimartingale and the non-semimartingale case. Finally, we review relevant findings for ambit processes in space-time.

\subsection{Volatility estimation for BSS processes}

An important class of ambit fields is obtained in the null-spatial or purely temporal case, often restricted to a (semi)-stationary setting.

In \cite{BNSch09},  the Brownian semistationary (BSS) process was introduced as follows. Let $(\Omega, \mathcal{F}, (\mathcal{F}(t))_{t\in \R}, \Prob)$ denote a probability space and  
$W$ an $(\mathcal{F}(t))_{t\in \R}$-adapted  Brownian motion on $\R$. The deterministic (kernel) function 
$g:\R \to \R$ is assumed to satisfy  $g(t)=0$ for $t\leq 0$ and $g\in L^2(\R)$,  and the $(\mathcal{F}(t))_{t\in \R}$-adapted c\`{a}dl\`{a}g stochastic volatility process is denoted by 
$\sigma=(\sigma(t))_{t\in \R}$. 
We shall assume that, 
for any $t \in \R$, 
\begin{align*}
\int_{-\infty}^t g^2(t-s) \sigma^2(s) ds < \infty, \quad \text{almost surely}.
\end{align*}
The BSS process (without drift term) is defined as $(Y(t))_{t\geq 0}$ with 
\begin{align}\label{eq:defBSS}
Y(t)=\int_{-\infty}^t g(t-s) \sigma(s) dW(s), \quad t \in \R. 
\end{align}

 A BSS process becomes a \Levy semistationary process if the Brownian motion $W$ is replaced by a general \Levy process $L$.

Following Ole's fruitful work on volatility estimation in semimartingale settings, highlighted in Neil Shephard's review, see \cite{Shephard2024}, Ole was interested in extending the ideas of realised variance and more general realised power variations to BSS processes. Related to this endavour is the question of under what conditions BSS processes are semimartingales.

Sufficient conditions for a BSS process to be a semimartingale were established in \cite{BNSch09} and we will state them briefly.
Suppose the following conditions hold:
(i) The function value $g(0+)$  exists and is finite;
(ii) the kernel function $g$ is absolutely continuous with square-integrable derivative $g'$;
(iii) the process $(g'(t-s)\sigma(s))_{s\in \R}$ is square integrable for each $t \in \R$.
Then $(Y(t))_{t\geq 0}$ is a semimartingale with representation 
\begin{align*}
Y(t) = Y(0) + g(0+) \int_0^t\sigma(s)dW(s) + \int_0^t a(s) ds, \qquad \text{for } t \geq 0,
\end{align*}
where, for $s \in \mathbb{R}$, we set  
\begin{align*}
a(s) &= \int_{-\infty}^s g'(s-u)\sigma(u)dW(u).
\end{align*}

We note that, in  the semimartingale case, the \BSS process can be expressed via   a stochastic differential equation (SDE), as follows 
\begin{align*}
 dY(t)= g(0+)\sigma(t)dW(t) + a(t)dt.
 \end{align*}

\subsubsection{The semimartingale case}
Consider the BSS process defined in \eqref{eq:defBSS}. In the special case when a  BSS process is a semimartingale, then the classic theory of volatility estimation via realised variance and power variation applies.

Suppose we observe the BSS process 
 over a time interval $[0, T]$ for $T>0$ at the observation times
$0, \Delta_n, 2 \Delta_n, \dots, \lfloor T/\Delta_n\rfloor \Delta_n$,
where $\Delta_n \to 0$ as $n \to \infty$ and $\lfloor \cdot \rfloor$ denotes the floor function.
The volatility estimation is carried out based on an 
 \emph{in-fill asymptotics} framework, 
where 
$\Delta_n \to 0$ and the length 
of the  observation interval $T$ is fixed.
Let us write
\begin{align*}
\Delta_i^n Y := Y(i \Delta_n)-Y((i-1)\Delta_n), \quad \text{ for } i = 1, \dots, \lfloor T/\Delta_n\rfloor,
\end{align*}
for 
the increments of the \BSS process $Y$. 

In the following, we will denote by 
 $\ucp$ uniform convergence on compacts in probability, see e.g.~\cite{Protter2005}, and by $\stackrel{d_{st}}{\longrightarrow }$
the concept of stable convergence in law, see e.g.~\cite{Renyi1963}.

The quadratic variation of $Y$ in $[0, T]$ is given by
\begin{align*}
[Y](t) = g(0+)^2 \int_0^t\sigma^2(s)ds, \qquad \text{for}\; t \in [0, T].
\end{align*}
The quadratic variation contains the accumulated squared volatility over the interval $[0,t]$. It can be estimated  consistently, using \emph{realised variance}, defined as 
\begin{align}\label{eq:RV}
\RV^n(t) = \sum_{i=1}^{\lfloor t/\Delta_n \rfloor}(\Delta_i^n Y)^2, \quad \text{for } t \in [0,T].
\end{align}

We will state a weak law of large numbers and a central limit theorem for realised variance. These were first derived in \cite{BNS2002, BNS2003, BNS2004} and later generalised in \citet[Theorems 2.4 and 2.9]{Jacod2008}.



\begin{theorem} \label{thm:main} Suppose that the volatility process $\sigma$ is \cadlag\ and adapted.
\begin{enumerate}  
\item[(i)] Law of large numbers: As $\Delta_n \to 0$,  \begin{align*}
\RV^n(t) \ucp g^2(0+)\int_0^t\sigma^2(s)ds, \quad \text{ as } \Delta_n \to 0. 
\end{align*}
\item[(ii)] Central limit theorem: 
As $\Delta_n \to 0$,  
\begin{align*}
\Delta_n^{-1/2}\left(\RV^n(t)-  g^2(0+)\int_0^t\sigma^2(s) ds \right)
\stackrel{d_{st}}{\longrightarrow }\sqrt{2} g^2(0+) \int_0^t \sigma^2(s)dB(s),
\end{align*}
on the Skorohod space $\mathcal{D}([0,T])$ equipped with the uniform topology. Here,  $B$ denotes a Brownian motion defined on an extension of the original probability space $(\Omega, \mathcal{F}, \Prob)$, which is independent of $\mathcal{F}$.
\end{enumerate}
\end{theorem}

Realised variance and extensions to other realised variation measures have been studied extensively in the last 25 years. For instance, jump-robust estimators have been developed by Ole and collaborators, see the work on realised multipower variation by \cite{BNGJPS2006} and \cite{BNSW2006} and also the survey by \cite{Shephard2024}. Around the same time, truncated realised variance was introduced, see \cite{Mancini2001}, \cite{Mancini2004}, \cite{Mancini2009}
and also \cite{Jacod2008}.
Some of these developments were further summarised in   \cite{JacodProtter2012} and  \cite{Ait-SahaliaJacod2014}.

\subsubsection{The non-semimartingale case}
Ole's work on ambit stochastics was greatly influenced by his interest in stochastic models for turbulence. Here, it turns out that the non-semimartingale case is of greater importance whereas the work on realised variation measures in a semimartingale setting was primarily driven by its potential for volatility estimation in finance.

As soon as we consider the non-semimartingale case, then the existence of a limit of realised variance is no longer guaranteed, and a new asymptotic theory is required. Ole and various collaborators embarked on the journey to establish this new asymptotic theory. 

An important insight in these theoretical developments was that the realised variance for BSS processes outside the semimartingale framework typically needs to be scaled appropriately in order to achieve convergence. The scaling factor typically depends on the kernel function of the BSS process.
Let us make this more precise in the following.

We define the \emph{Gaussian core} (GC) as a special case of a \BSS process with $\sigma\equiv 1$ given by  
\begin{align*}
\mathrm{GC}(t) := \int_{-\infty}^t g(t-s)\, dW(s), \quad t \in \R,
\end{align*}
with autocorrelation function  denoted  by 
\begin{align*}
r(t) = \text{Cor}(\mathrm{GC}(u), \mathrm{GC}(u+t)) = \frac{\int_0^{\infty}g(s)g(s+t) ds}{||g||^2_{L^2(\R)}}, \quad t \geq 0,
\end{align*}
where $||g||^2_{L^2(\R)}=\int_0^{\infty}g^2(s)ds$.
The variogram of the Gaussian core is given by 
\begin{align*}
R(t) := \EE[(\mathrm{GC}(t+u)-\mathrm{GC}(u))^2] =2||g||^2_{L^2(\R)} (1-r(t)). 
\end{align*}
We define the scaling constant
\begin{align*}
\tau_n:= \tau(\Delta_n) := \sqrt{\EE[(\Delta_{i}^n \mathrm{GC})^2]}=\sqrt{R(\Delta_n)}=\sqrt{2||g||^2_{L^2(\R)} (1-r(\Delta_n))}.
\end{align*}


 We summarise the asymptotic  theory presented in \citet[Theorems 3.1 and 3.2]{CorcueraHPP2013}, see also \cite{Podolskij2014,BNCP11, BNCP10b}.
 In the following assumption, all functions $L_f:(0,\infty)\to \R$, indexed by a mapping $f$, are assumed to be continuous and slowly varying at 0. Also, $f^{(m)}$ denotes the $m$th derivative of the function $f$. 
 The following assumption were formulated in  \citet[Assumptions A1, A2] 
 {CorcueraHPP2013}, who consider  the case when $g$ has a single singularity at $0$, the case of multiple singularities was considered in \cite{Gaertner2015}.
 \begin{assumption}\label{assumption:Cha3_A}
Suppose that 
there exists an $\alpha\in (-\frac{1}{2},0)\cup(0, \frac{1}{2})$ such that  the following conditions hold :
\begin{enumerate}
\item[(i)] $g(x)=x^{\alpha}L_g(x)$. 
\item[(ii)] $g'(x)=g^{(1)}(x)=x^{\alpha-1}L_{g'}(x)$ and $g'\in L^2((\epsilon,\infty))$, for any $\epsilon>0$. Also, there exists a constant $a>0$ such that $g'$ is non-increasing on $(a, \infty)$. 
\item[(iii)] 
For any strictly positive $t$, we have that 
\begin{align*}
F(t) = \int_{1}^{\infty}(g'(s))^2 \sigma^2(t-s) ds < \infty.
\end{align*}
\item[(iv)] The variogram of the Gaussian core satisfies $R(x)=x^{2\alpha+1}L_R(x)$.
\item[(v)] $R^{(2)}(x)=x^{2\alpha-1}f_{R^{(2)}}(x)$.
\item[(vi)] There exists a constant $b\in (0,1)$ such that 
$\limsup_{x\downarrow 0} \sup_{y\in [x,x^b]}\left|\frac{L_{R^{(2)}}(y)}{L_R(x)}\right|<\infty$.
\end{enumerate}
\end{assumption}

We have the following results,  cf.~\citet[Theorems 1 and 2]{BNCP10b}.
\begin{theorem}\label{th:Cha3_CLT}
Suppose that Assumption \ref{assumption:Cha3_A} holds.
\begin{enumerate}
\item[(i)]
Law of large numbers: As $\Delta_n \to 0$, we have
\begin{align*}
\Delta_n \tau_n^{-2} \RV^n(t) 
\ucp  \int_0^t\sigma^2(s) ds.
\end{align*}
\item[(ii)] Central limit theorem: Suppose that the volatility process $\sigma$ is H\"{o}lder continuous of order $1/2<\gamma <1$  and suppose  that $\alpha \in (-1/2,0)$. Then 
\begin{align*}
\Delta_n^{-1/2}\left(\Delta_n \tau_n^{-2}\RV^n(t) -\int_0^t\sigma^2(s) ds\right) \stackrel{d_{st}}{\longrightarrow } K\int_0^t \sigma^2(s) dB(s),
\end{align*}
on the Skorohod space $\mathcal{D}([0,T])$ equipped with the uniform topology, where $B$ denotes a Brownian motion defined on an extension of the original probability space $(\Omega, \mathcal{F}, \Prob)$, which is independent of $\mathcal{F}$. Also
\begin{align*}
K^2 := \lim_{n \to \infty}\Delta_n^{-1}\Var\left(\Delta_n^{1-2H}
\sum_{i=1}^{\lfloor 1/\Delta_n\rfloor}(\Delta_i^n B^H)^2\right),
\end{align*}
where $B^H$ denotes a fractional Brownian motion with Hurst parameter $H = \alpha + 1/2$. 
\end{enumerate}
\end{theorem}
Although the above result is interesting from a mathematical point of view, it is infeasible since the scaling factor $\tau_n$ is unknown in practice. 
This problem can be avoided by considering suitable ratio statistics. Let us consider the setting of a \BSS process with a \emph{gamma kernel}, as studied in 
 \cite{BNSch09}.  
 The gamma kernel is proportional to 
\begin{align}\label{eq:Gammakernel}
g(x) = x^{\alpha}e^{-\lambda x}, \quad \text{ for } \lambda > 0,\ \alpha \in (-1/2,0)\cup (0,1/2), \; \mathrm{for} \; x>0,
\end{align}
and $g(x)=0$ for $x\leq 0$
and satisfies 
 Assumption \ref{assumption:Cha3_A}. 
Hence, the above asymptotic theory is applicable. 
Moreover, we can consider the ratio of two power variations evaluated over different frequencies termed the \emph{change-of-frequency statistics}, see \citet[Section 4]{CorcueraHPP2013}:
\begin{corollary} For fixed $t>0$ and $g$ given by \eqref{eq:Gammakernel}, we have
\begin{align*}
S_n (t):=
\frac{\sum_{i=1}^{\lfloor t/\Delta_n \rfloor}(Y(i\Delta_n)-Y((i-2)\Delta_n))^2}{\sum_{i=1}^{\lfloor t/(2\Delta_n) \rfloor}(Y(i\Delta_n) - Y((i-1)\Delta_n))^2} \stackrel{\Prob}{\longrightarrow }2^{2\alpha + 1}, \quad \text{ as } \Delta_n \to 0,
\end{align*}
and, hence, 
\begin{align*}
\widehat \alpha_n = \frac{1}{2}\left(\log_2 S_n(t)-1 \right) \stackrel{\Prob}{\longrightarrow } \alpha, \quad \text{ as } \Delta_n \to 0, 
\end{align*}
where $\log_2$ denotes the logarithm to base 2 and $\stackrel{\Prob}{\longrightarrow }$ denotes convergence in probability.
\end{corollary}
\cite{BNCP11, BNCP10b,CorcueraHPP2013}  extended the above result further and also developed a suitable central limit theorem. 

While estimating integrated volatility requires some knowledge of $\tau_n$ (or a suitable estimator thereof), 
\emph{relative volatility} can  be estimated without further knowledge of the scaling factor, see \cite{BNPS2015}.  
Consider a fixed  time $T>0$ and  define the \emph{relative volatility} by 
\begin{align*}
\mathrm{RelVol}(t) := \frac{\int_0^t \sigma^2(s)ds}{\int_0^T \sigma^2(s)ds}, \quad \text{ for } t \leq T.
\end{align*}
\begin{corollary} \begin{enumerate}
\item Law of large numbers: Under the same assumptions as in Theorem \ref{th:Cha3_CLT}, we have that
\begin{align*}
\widehat{\mathrm{RelVol}^n(t)}:=\frac{\RV^n(t)}{\RV^n(T)} \stackrel{\Prob}{\longrightarrow } \mathrm{RelVol}(t), \quad \text{ as } \Delta_n \to 0,
\end{align*}
uniformly in $t\in [0,T]$.
\item Central limit theorem: 
Assume the same H\"older regularity of the volatility process $\sigma$ as described in Theorem \ref{th:Cha3_CLT}(ii), then:
\begin{multline*}
\Delta_n^{-1/2}\left(\widehat{\mathrm{RelVol}^n(t)} -\mathrm{RelVol}(t)\right)\\
\stackrel{d_{st}}{\longrightarrow } \frac{K}{\int_0^T\sigma^2(s)ds} \left(\int_0^t \sigma^2(s) dB(s)-\mathrm{RelVol}(t)\int_0^T \sigma^2(s) dB(s)\right),
\end{multline*}
on the Skorohod space $\mathcal{D}([0,T])$ equipped with the uniform topology, where $B$ denotes a Brownian motion defined on an extension of the original probability space $(\Omega, \mathcal{F}, \Prob)$, which is independent of $\mathcal{F}$. 
\end{enumerate}
\end{corollary}

We  presented the results above for realised variance to simplify the exposition. However, similar results hold for general $p$-power variations as outlined in the articles mentioned throughout the section and also in \cite[Chapter 3]{BNBV2018_ambitbook}. We also note that an additional  drift term can often be included without impacting the presented asymptotic theory. We refer to \cite{BNCP11} for a detailed study of this aspect. 

While the results in the semimartingale and non-semimartingale \enquote{look}  similar apart from the corresponding scaling terms in the realised variance, it is important to note that the method of the proofs varies significantly in the two cases. For the non-semimartingale case, the proofs frequently proceed as follows: 
  We typically focus on the  Gaussian core first, i.e.~a Brownian semistationary process without stochastic volatility and derive all results in the absence of stochastic volatility. Here, we need to find a suitable scaling factor for the realised variance.
 We then extend all results to the general case with stochastic volatility using Bernstein's blocking technique, where the volatility is \enquote{frozen}  on a coarser time grid.
The weak law of large numbers can be proven using classical techniques for the convergence of measures.
  For the central limit theorem, we typically invoke the powerful fourth-moment theorem by \cite{NP2005}.  

\subsection{Volatility estimation for ambit processes}
The third scenario in volatility estimation we will discuss in this survey, based on \cite{BNG2011}, is within a spatio-temporal setting, where the literature is much less developed than in the temporal case. 

Consider an ambit field in the 1-dimensional spatial case with Gaussian \Levy basis $W$, i.e.~
\begin{equation*}
Y(x,t)=\int_{A(x,t) }g(x - \xi, t-s)\sigma( \xi, s)W(d\xi, ds),
\end{equation*}
with an ambit set 
satisfying $A(x,t)=A+(x, t)$, for some  $A:=A(0,0)\subseteq \R^2$. In the following, we will assume that $\sigma$ is a continuous random field, independent of $W$, with $(\xi, s)\mapsto \EE(\sigma^2(\xi,s))$ being locally bounded. 

Let
$v(\theta)=(x(\theta), t(\theta))$ denote a smooth curve in space-time, where $\theta\mapsto t(\theta)$ is nondecreasing.
 Define
$$
X(\theta)=Y(x(\theta), t(\theta)),
$$
and assume that $X=(X(\theta))_{\theta \in \R}$ is a well-defined stochastic process, which we will call an \emph{ambit process}. 
Some natural questions we might ask are: When is the quadratic variation of $X$ well defined?
 Can we define a stochastic differential $dX(\theta)$?
 In the situation where $X$ is a semimartingale, the answer is \enquote{yes} to the two above questions. Outside the semimartingale framework, the situation is more complex, but some of the insights formulated in the purely temporal case carry over, in particular, when it comes to scaling the realised variance appropriately to obtain convergence.

We review the key results by  
\cite{BNG2011} in the following. They  were particularly interested in the case of a closed set
\begin{align*}
    A=\{(\xi, s)\in \R^2| -M\leq s\leq 0, c_1(s)\leq \xi \leq c_2(s)\},
\end{align*}
for a constant $M>0$ that can be considered as decorrelation time and two smooth functions $c_1:[-M,0]\to(-\infty,0)$ and $c_2:[-M,0]\to(0,\infty)$, where $c_1$ is increasing and $c_2$ decreasing.

As before, one can then define the realised variance as 
\begin{align*}
\RV^n(t) = \sum_{i=1}^{\lfloor t/\Delta_n \rfloor}(\Delta_i^n X)^2
=\sum_{i=1}^{\lfloor t/\Delta_n \rfloor}(X(i\Delta_n)-X((i-1)\Delta_n))^2, \quad \text{for } t > 0,
\end{align*}
and the scaled realised variance as 
\begin{align*}
\SRV^n(t) = \frac{\Delta_n}{c(\Delta_n)}\RV^n(t), \quad \text{for } t > 0,
\end{align*}
for a positive scaling function $c$. 

The key result derived in \cite{BNG2011} is the following: Under suitable technical conditions, we have that
\begin{align}\label{eq:ambitconv}
\SRV^n(t)\stackrel{\Prob}{\to} \int_{\R^2}\int_0^t\sigma^2(x(s)-u, t(s)-u')ds \pi(du, du'), \qquad \mathrm{as}\; \Delta_n \to 0,
\end{align}
where the measure $\pi$ is constructed below. 

\subsubsection{Technical conditions for the weak law of scaled realised variance}
Let us briefly summarise the (additional) technical conditions used by \cite{BNG2011} to establish the weak law of scaled realised variance stated in \eqref{eq:ambitconv}. 

The smooth curve $v$ is assumed to be a straight line such that 
\begin{align*}
    \Delta v(\Delta_n):=(\Delta x(\Delta_n), \Delta t(\Delta_n)):=(x(z+\Delta_n)-x(z), t(z+\Delta_n)-t(z)), 
\end{align*}
for $z, \Delta_n >0$. 

In order to define the probability measure $\pi$ above, we first introduce the function $\psi$ as follows:
\begin{align*}
&\psi_{\Delta_n}(u,u')\\
&:=\left \{
\begin{array}{ll}
(g(\Delta x(\Delta_n)+u, \Delta t(\Delta_n)+u')-g(u,u'))^2, & \mathrm{for}\; (u,u')\in (-A)\cap (-A-\Delta v(\Delta_n)), \\
g^2(u,u'), & \mathrm{for}\; (u,u')\in (-A)\setminus (-A-\Delta v(\Delta_n)), \\
g^2(\Delta x(\Delta_n)+u, \Delta t(\Delta_n)+u'), & \mathrm{for}\; (u,u')\in (-A-\Delta v(\Delta_n))\setminus (-A).
\end{array}
\right.
\end{align*}
From the above definition, we can see that $\psi_{\Delta_n}(u,u')=0$ for $(u,u')\not \in (-A)\cup (-A-\Delta v(\Delta_n))$ and that, under suitable smoothness conditions, $\psi_{\Delta_n}(u,u')$ typically has order $\Delta_n^2$ for $(u,u') \in (-A)\cap (-A-\Delta v(\Delta_n))$.
The scaling function $c$ can now be defined as
\begin{align*}
    c(\Delta_n):=\int_{\R^2}\psi_{\Delta_n}(u,u')du du',
\end{align*}
assuming $c(\Delta_n)>0$, which will be the case under the additional assumptions given below. 
One can then define the probability measure, for $\Delta_n>0$, as  
\begin{align*}
    \pi_{\Delta_n}(du,du'):=\frac{\psi_{\Delta_n}(u,u')}{c(\Delta_n)}du du'.
\end{align*}
Note that any weak limit point $\pi_0$ of $\pi_{\Delta_n}$ is a probability measure concentrated on $(-A)$. We then define $\pi$ as the image measure of $\pi_0$ under the transformation $(u,u')\mapsto(-u, -u')$ and note that it is concentrated on $\partial A$.

We can now state the key convergence result, see \cite[Theorem 1]{BNG2011}:
\begin{theorem}
Let $A$ be a bounded, closed, convex set with nonempty interior $\interior{A}$ and piecewise $C^{\infty}$ boundary. Suppose that $v$ is a straight line and suppose that, for some $-1/2<\alpha<1/2$, the kernel function $g$ can be represented as $g({\bf z})=\phi({\bf z})h_{\alpha}({\bf z})$, where $\phi$ is Lipschitz continuous and not identically 0 of the part of $\-\partial A$ nonparallel to $v$, and  
\begin{align*}
    h_{\alpha}(-{\bf z})=\left \{
\begin{array}{ll}
(1-T({\bf z}-{\bf z}_0))^{\alpha}, & \mathbf{z} \in A,
\\
0, & \mathbf{z} \not \in A,
\end{array}
    \right.
\end{align*}
where the function $T$ denotes the gauge function of the set $A-{\bf z}_0$ for some ${\bf z}_0\in \interior{A}$.
Then there exists a probability measure $\pi$ concentrated on $\partial A$ such that \eqref{eq:ambitconv}  holds. 
\end{theorem}
\begin{remark}
Gauge functions are discussed in 
\citet[Chapter 5]{DS1958}: Let $C\in \R^2$ denote a bounded, closed, convex set with ${\bf 0}\in \interior{C}$. Then $T({\bf z}):=\inf\{t>0|{\bf z}\in tC\},$ for ${\bf z}\in \R^2$ is called the gauge function of $C$.
\end{remark}

\subsection{Further reading}
In the last two decades, research on limit theorems for realised variation measures and related quantities has developed at a fast pace. Early works in the semimartingale context were linked to important applications in financial econometrics, see \cite{Shephard2024} and, amongst others, \cite{BNGJPS2006,BNGJS2006,BNSW2006}. Moreover,  \cite{Ait-SahaliaJacod2014}  give a comprehensive overview of the corresponding results up to 2014.

In the non-semimartingale case, new tools based on Malliavin calculus and the 4th-moment theorem, see \cite{NP2005}, enabled a second wave of research papers in this realm.
For example, work on realised power variation for Gaussian, Brownian and \Levy stationary processes include, in the univariate case, 
\cite{BNCPW2009, 
BNCP11,  
Corcueraetal2013,CorcueraHPP2013,Gaertner2015,BLP2017, 
BasseOConnorHeinrichPodolskij2016},
and, in the multivariate case, \cite{GranelliVeraart2019,PasseggeriVeraart2019,LPV2023}.

In the spatial setting, limit theorems for power variations of ambit fields have been established in 
\cite{Pakkanen2014}; \cite{NguyenVeraart2016}
study spatial heteroskedasticity in volatility modulated moving averages.
Recently, the limit theory for realised variance has been extended to infinite-dimensional settings, see 
\cite{BenthSchroersVeraart2022, BenthSchroersVeraart2024}, which we will briefly describe next in Section \ref{sec:SPDElinks}.

\section{Ambit fields and SPDEs}\label{sec:SPDElinks}

Ambit fields $Y(\mathbf{ x},t)$ in \eqref{eq:Cha0_ambitfield} are stochastic models of the evolution of physical phenomena in time and space. Usually, in science, dynamic systems are modelled by partial differential equations accounting for random fluctuations by including stochastic terms. Ambit fields provide a new perspective on such stochastic partial differential equations, allowing for more flexibility in the modelling of the probabilistic features of the phenomenon we want to describe. We discuss this in the following,
referring the interested reader to \cite{BNBV2011} for a more in-depth analysis. 

Following \cite{PeszatZabczyk}, let $U$ be a L\'evy process taking values in the separable Hilbert space $H$ with covariance operator $Q$ being positive definite and trace class. Further, let $(\mathcal S(t))_{t\geq 0}$ be a $C_0$-semigroup with a densely defined generator $\mathcal A$ on $H$, and consider the  class of linear parabolic SPDEs given as
\begin{equation}
\label{parabolic-spde}
 dZ(t)=\mathcal A Z(t) dt+\nu(t) dU(t),
\end{equation}
with $Z(0)=Z_0\in H$ given and $(\nu(t))_{t\geq 0}$ being a predictable process taking values in the space of linear operators on $H$ such that $\mathbb E[\int_0^T\Vert\nu(s) Q^{1/2}\Vert^2_{\text{HS}}\,ds]<\infty$ for $T<\infty$. A mild solution is a predictable $H$-valued process
$(Z(t))_{t\in[0,T]}$ such that 
\begin{equation}
\label{mild-parabolic-spde}
  Z(t)=\mathcal S(t)Z_0+\int_0^t\mathcal S(t-s)\nu(s)dU(s).
\end{equation}
Often, $H$ is some space of real-valued functions on $\mathbb R^d$ (or a subset thereof). For example, $H$ could be $L^2(\mathbb R^d)$, some
Sobolev space on $\mathbb R^d$, or the Filipovic space on $\mathbb R_+$ (see \cite{BNBV2018_ambitbook}).  Furthermore, one can  express the semigroup operator $\mathcal S(t)$ as an integral operator in terms of a Green's function $\mathcal S(t)f(\mathbf x)=\int_{\mathbb R^d}g(\mathbf x,t-s;\boldsymbol{\xi})f(\boldsymbol{\xi})d\boldsymbol{\xi}$. Moreover, the $H$-valued L\'evy process $U$ may be rephrased into a L\'evy basis $L$. If $H$ is a Banach algebra, as
is the case for the Filipovic space, or imposing additional assumptions on the process $\nu$, we may express $\nu(t)dU(t)$ as a
pointwise multiplication operator $\sigma(\boldsymbol{\xi},t)L(d\boldsymbol{\xi},dt)$ for $(\sigma(t))_{t\geq 0}$ being a predictable $H$-valued process accounting for $\nu$ and the covariance structure of $U$ encoded in the $Q$-operator. After some reformulations based partly on the theory of \cite{W}, we can express the mild solution in \eqref{mild-parabolic-spde} as
\begin{equation}
\label{SPDE-ambit-field}
 Z(\mathbf{x},t)=\int_0^t\int_{\mathbb R^d}g(\mathbf{x}, t-s;\boldsymbol{\xi})\sigma(\boldsymbol{\xi},s)L(d\boldsymbol{\xi},ds).
\end{equation}
For simplicity, we let $Z_0=0$. Notice that $\mathbf x\mapsto Z(\mathbf{x},t)$ is a an element in the chosen Hilbert space of functions. Indeed, we see that $Z(\mathbf{x},t)$ is an ambit field
as defined in \eqref{eq:Cha0_ambitfield}. 

We remark in passing that ambit fields are Volterra processes in time, and can be lifted to be mild solutions of parabolic SPDEs where the generator is given by $\mathcal A=\nabla_{\mathbf x}$. We refer to \cite{BE2015} for further details.

Worth observing here is that the kernel function $g$ appears as a Green's function of the operator $\mathcal A$. In real-world 
applications, the operator $\mathcal A$ may not be precisely known. Also, it may be difficult to analytically express the Green's function. In the perspective of ambit fields, $g$ models the dependency structure in space and time, and offers a significant extension in modelling flexibility beyond Green's functions, where it may even be an approximation based on probabilistic considerations. Furthermore, $g$ can also be re-phrased accounting for an ambit set moving in time and space, further enhancing the flexibility in describing the time-space evolution of the phenomenon in question.  
The connection with mild solutions of parabolic SPDEs opens, on the other hand, a way on dealing with realized variance for ambit fields.

\subsection{Volatility estimation for SPDEs}
 In \cite{BenthSchroersVeraart2022,BenthSchroersVeraart2024}, a law of large numbers and
central limit theorem are presented for the volatility 
$\nu$ in parabolic SPDEs \eqref{parabolic-spde} driven by an $H$-valued {\it Wiener} process $U$.

Having an equally spaced time grid $t_i=i\Delta_n$ for $\Delta_n=1/n$ and $i=1,2,\ldots, \lfloor t/\Delta_n\rfloor$, \cite{BenthSchroersVeraart2022} introduced the {\it semigroup-adjusted realized covariation} (SARCV) as
\begin{equation*}
    \sum_{i=1}^{\lfloor t/\Delta_n\rfloor} (Z(t_i)-\mathcal S(\Delta_n)Z(t_{i-1}))^{\otimes 2},
\end{equation*}
 where $\otimes$ denotes the tensor product in the Hilbert space $H$ (note here that we are back to a general separable Hilbert space). The SARC defines a sequence of stochastic processes on $[0,T]$ taking values in the space of Hilbert-Schmidt valued operators on $H$.  In \cite{BenthSchroersVeraart2022}, it is shown that the SARCV converges uniformly on compacts in probability (ucp) in the Hilbert-Schmidt norm to the integrated covariance process, i.e., for all $\epsilon>0$ and every $T>0$,  
\begin{equation*}   \lim_{n\rightarrow\infty}\mathbb P\left(\sup_{0\leq t\leq T}\left\|\sum_{i=1}^{\lfloor t/\Delta_n\rfloor}(Z(t_i)-\mathcal S(\Delta_n)Z(t_{i-1}))^{\otimes 2}-\int_0^t\nu(s)Q\nu^*(s)ds\right\|_{\text{HS}}>\epsilon\right)=0
\end{equation*}
 under rather weak conditions. Here, $\nu^*(s)$ is the adjoint of $\nu(s)$.  

Under a fourth moment integrability condition on the Hilbert-Schmidt norm of the volatility process $\nu$ one can show central limit theorems for the SARCV. To describe these, we introduce the operator-valued process $(\Gamma(t))_{t\geq 0}$ taking values in the space of Hilbert-Schmidt operators on $H$ into itself, denoted $\mathcal H:=L_{\text{HS}}(H,H)$: for $B\in\mathcal H$, 
\begin{equation*}
    \Gamma(t) B=\int_0^t(\nu(s) Q\nu^*(s))(B+B^*)(\nu(s) Q\nu^*(s))ds.
\end{equation*}
If the semigroup $\mathcal S$ satisfies a certain regularity condition after acting on the volatility process, we have that the stochastic process 
$$
\left(\Delta_n^{-1/2}\left(\sum_{i=1}^{\lfloor t/\Delta_n\rfloor}(Z(t_i)-\mathcal S(\Delta_n)Z(t_{i-1}))^{\otimes 2}-\int_0^t\nu(s)Q\nu^*(s)ds\right)\right)_{t\in[0,T]}
$$
converges, when $n\rightarrow\infty$ stably in law in the Skorohod space $\mathcal D([0,T],\mathcal H)$
to a conditionally continuous Gaussian $\mathcal H$-valued independent increment process with mean zero and covariance operator process $\Gamma$. One could dispense on the regularity condition on the semigroup, and achieve a central limit theorem for finite-dimensional projections of the SARCV. Indeed, letting 
$B=\sum_{j=1}^kc_j h_j\otimes g_j$ for $c_j\in\mathbb R$ and $h_j,g_j$ belonging to a certain subspace of $H$, we have that 
$$
\left(\Delta_n^{-1/2}\left\langle\left(\sum_{i=1}^{\lfloor t/\Delta_n\rfloor}(Z(t_i)-\mathcal S(\Delta_n)Z(t_{i-1}))^{\otimes 2}-\int_0^t\nu(s)Q\nu^*(s)ds\right),B\right\rangle_{\mathcal H}\right)_{t\in[0,T]}
$$
converges stably in law in the Skorohod space $\mathcal D([0,T],H)$ to a conditionally continuous
Gaussian $H$-valued independent increment process with mean zero and covariance operator process $\langle\Gamma B,B\rangle_{\mathcal H}$. The elements $h_j,g_j$ belong to the so-called $1/2$-Favard space defined by the adjoint of the semigroup. In the finite-dimensional central limit theorem, we have moved the regularity from the semigroup $\mathcal S$ to a class of finite-dimensional operators $B$ on which we can project the SARCV.   

In practice, we do not know the asymptotic covariance operator
$\Gamma$ and hence the central limit theorems are infeasible. In order to have feasible limit theorems, we introduce an estimator of $\Gamma$ using semigroup-adjusted multipower variation (SAMPV). In particular, one uses the fourth-power and second-bipower variation processes, defined as
\begin{equation*}    SAMPV^n(t;4)=\sum_{i=1}^{\lfloor t/\Delta_n\rfloor}(Z(t_i)-\mathcal S(\Delta_n)Z(t_{i-1}))^{\otimes 4},
\end{equation*}
and
\begin{equation*}
   SAMPV^n(t;2,2)= \sum_{i=1}^{\lfloor t/\Delta_n\rfloor-1}(Z(t_i)-\mathcal S(\Delta_n)Z(t_{i-1}))^{\otimes 2}\otimes(Z(t_{i+1})-\mathcal S(\Delta_n)Z(t_{i}))^{\otimes 2},
\end{equation*}
respectively. Define the estimator
\begin{equation*}
\widehat{\Gamma}^n(t)=\Delta_n^{-1}\left(SAMPV^n(t;4)-SAMPV^n(t;2,2)\right).
\end{equation*}
This defines a consistent estimator of $\Gamma$, since
$\widehat{\Gamma}^n\rightarrow\Gamma$ when $n\rightarrow\infty$ in $\mathcal H^4$ uniformly on compacts in probability (ucp). Here, $\mathcal H^4=L_{\text{HS}}(H,L_{\text{HS}}(H,\mathcal H))$. If the volatility process $\nu$ is c\'adl\'ag in the space of Hilbert-Schmidt operators, we obtain a feasible central limit theorem: As $n\rightarrow\infty$,  
$$
\Delta_n^{-1/2}\langle\widehat{\Gamma}^n(t)B,B\rangle_{\mathcal H}^{-1/2}\left\langle\left(\sum_{i=1}^{\lfloor t/\Delta_n\rfloor}(Z(t_i)-\mathcal S(\Delta_n)Z(t_{i-1}))^{\otimes 2}-\int_0^t\nu(s)Q\nu^*(s)ds\right),B\right\rangle_{\mathcal H},
$$
converges in distribution to a standard normal random variable. As above, $B$ is a finite-dimensional operator defined as tensors of elements in the $1/2$-Favard subspace of $H$. We refer to \cite{BenthSchroersVeraart2024} for details on the above discussion of central limit theorems. 

\subsection{Links to turbulence}
Let us return to parabolic SPDEs.  \cite{PeszatZabczyk} study nonlinear parabolic SPDEs driven by L\'evy noise defined on a Hilbert space $H$, of the form 
\begin{equation}
\label{SPDE-general}
    dZ(t)=(\mathcal A+F(Z(t)))dt+G(Z(t))dU(t),
\end{equation}
where $F$ is a Lipschitz map from $H$ into itself while $G$ is a Lipschitz map from $H$ into Hilbert-Schmidt operators, both of at most linear growth in appropriate norms (see \cite{PeszatZabczyk} for precise conditions). A mild solution is defined as the process $Z$ solving the nonlinear integral equation
\begin{equation}
\label{mild-nonlin-spde}
    Z(t)=\mathcal S(t) Z_0+\int_0^t\mathcal S(t-s)F(Z(s))ds+\int_0^t\mathcal S(t-s)G(Z(s))dU(s),
\end{equation}
where $Z(0)=Z_0$ is the initial condition (again we refer to \cite{PeszatZabczyk} for additional regularity conditions).
\cite{Bir13b} studies the stochastic Navier-Stokes equation for fully developed turbulence within this context. In his formulation, supposing that the fluid satisfies periodic boundary conditions on its domain, the motion $u=(u(\mathbf{x},t))_{\mathbf{x}\in\mathbb T^3,t\geq 0}$ of the fluid follows the nonlinear parabolic SPDE dynamics
\begin{align*}
    du(\mathbf{x},t)&=(\nu\Delta u(\mathbf{x},t)-u(\mathbf{x},t)\cdot\nabla u(\mathbf{x},t)+\nabla\Delta^{-1}\text{tr}(\nabla u(\mathbf{x},t))^2)dt+\sum_{\mathbf{k}\in\mathbb Z^3}c_{\mathbf{k}}^{1/2}db^{\mathbf{k}}(t)e_{\mathbf{k}}(\mathbf{x}) 
    \\
    &\qquad+\sum_{\mathbf{k}\in\mathbb Z^3, \mathbf{k}\neq 0}d_{\mathbf{k}}\eta_{\mathbf{k}}dt e_{\mathbf{k}}(\mathbf{x})+u(\mathbf{x},t)\sum_{\mathbf{k}\in\mathbb Z^3,\vert \mathbf{k}\vert\leq m,\mathbf{k}\neq 0}\int_{\mathbb R}h_{\mathbf{k}}\overline{N}^{\mathbf{k}}(dt,dz),
\end{align*}
where $u(\cdot,t)\in L^2(\mathbb T^3)$. Here, $e_{\mathbf{k}}(\mathbf{x})=\exp(2\pi\mathrm i \mathbf{k}\cdot \mathbf{x})$ are the Fourier coefficients, and $b^{\mathbf{k}}$ are independent Brownian motions. Furthermore, we have finitely many jump processes modelled by the compensated Poisson measures $\overline{N}^{\mathbf{k}}$, $\vert\mathbf{k}\vert\leq m$.    

The equations for the invariant measures of both the one-point and two-point statistics of turbulence are computed. The two-point statistics describes the velocity differences $u(\mathbf{x},t)-u(\mathbf{y},t)$, and \cite{Bir13b} provides an explicit solution of the PDE describing the probability density function of the invariant measure. It turns out to be a normal inverse Gaussian distribution (NIG), corresponding to empirical turbulence data as evidenced in \cite{BNBlSch04} (see also \cite{Bir13b}). 

Rather than studying integral equations such as the mild formulation \eqref{mild-nonlin-spde} of the SPDE \eqref{SPDE-general}, one can define Volterra equations where the semigroup operator $\mathcal S$ is substituted with some integral kernel. This is analogous to extending the mild solution of a linear parabolic SPDE to ambit fields \eqref{SPDE-ambit-field}. \cite{Chong2017} provides the existence and uniqueness of solutions to such equations, being on the form
\begin{equation*}
    Z(\mathbf{x},t)=Z_0(\mathbf{x})+\int_0^t\int_{\mathbb R^d} g(\mathbf{x},t;\boldsymbol{\xi},s)\sigma(Z(\boldsymbol{\xi},s))L(d\boldsymbol{\xi},ds).
\end{equation*}
Here, the stochastic volatility is state-dependent, and the deterministic function $\sigma$ is supposed to be Lipschitz continuous. In the null-space case, similar 
Volterra equations have been lifted to parabolic SPDEs and analysed in \cite{BenthDetKruhner2022}. 

\subsection{Further reading}

There is an interesting relationship between ambit fields and SPDEs formulated within the White Noise Analysis context where the derivative of the noise is the basis object, see \cite{BNBV2011} for details. For a general theory on SPDEs studied from the White Noise Analysis viewpoint, we refer to \cite{HOUZ}. Cylindrical Wiener and L\'evy processes play a central role in the formulation and analysis of SPDEs. \cite{AppleRiedle} define cylindrical L\'evy processes in Banach spaces. In \cite{Apple} a survey on infinite dimensional (Hilbert space-valued) Ornstein-Uhlenbeck processes driven by L\'evy processes is given. The linear parabolic SPDEs discussed above are naturally viewed as infinite-dimensional Ornstein-Uhlenbeck processes. Such SPDEs are the modelling tool for commodity forward curves, see \cite{BenthKruhner}, but ambit fields provide an attractive and flexible extension, see \cite{BNBV2018_ambitbook}. In \cite{BS_meta} the relationship between metatimes, cylindrical random variables and random measures is studied.

\section{Ambit stochastics in practice}\label{sec:practice}
Ambit fields can be applied in many fields. In the following, we illustrate their potential in turbulence and growth models, the first application areas considered in the literature on ambit stochastics, see \cite{BNSch07a}.
Later, L\'{e}vy/Brownian semistationary processes and ambit fields have been used in other applications, including the  
stochastic modelling of energy markets, see e.g.~\cite{BNBV2013, Bennedsen2015}, weather derivatives and temperature, see \cite{BB2009, BenthBenth2011, BenthBenth2012book},  short rates, see \cite{Corcueraetal2013},  
environmental applications, see e.g.~\cite{NguyenVeraart2016} and financial applications, such as  rough volatility models, see \cite{BLP2021}, and high-frequency financial data, see \cite{BLSV2023}.

\subsection{Turbulence}

Throughout his research career, Ole focused extensively on
stochastic models and statistical inference techniques that
are applicable to physical turbulence; this area has been
at the heart of advances in ambit stochastics ever since.
It is important to ensure that such stochastic models for turbulence align with the \emph{Statistical Theory of
Turbulence}  founded by Kolmogorov and Obukhov, see \cite{Frisch95,Bir13b} for surveys. 

 Based on the detailed summary in    
\citet[Chapter 9]{BNBV2018_ambitbook},  we will provide a short overview of some of the key aspects at the interface of turbulence and ambit stochastics.

Turbulence research typically focusses on the dynamics of the three-dimensional  velocity vector $\mathbf{v}(\mathbf{x},t)=(v_{x}(\mathbf{x}
,t),v_{y}(\mathbf{x},t),v_{z}(\mathbf{x},t))$ as a function of position
$\mathbf{x}
=(x,y,z)$ and time $t$ in a fluid flow. 
One can then define the 
  \emph{energy dissipation} in an incompressible flow 
as
\begin{align} 
  \label{epsfull}
  \varepsilon(\mathbf{x},t):= \frac{\nu}{2}\sum_{i,j=x,y,z}
  \left(\partial_{i}v_{j}(\mathbf{x},t)+\partial_{j}v_{i}(\mathbf{x},t)\right)^{2},
\end{align}
which 
represents the loss of kinetic energy due to friction
forces. The scaling factor $\nu$ denotes the viscosity, measuring the resistance to gradual deformation by shear stress or tensile stress. 

When carrying out high-resolution experiments, one typically obtains measurements of a time series of a single component of the velocity vector in the direction of the mean flow and at a single location  $\mathbf{x}_{0}$.
Therefore, it is common to define the \emph{temporal energy dissipation} for stationary, homogeneous and
isotropic flows as 
\begin{align}
  \label{epssurr}
   \widetilde{\varepsilon}(\mathbf{x}_{0},t)
  \equiv \frac{15\nu}{\overline{v}^{2}}
  \left(\frac{dv(\mathbf{x}_{0},t)}{dt}\right)^{2}\,,
\end{align}
where the constant $\overline{v}$ represents the  mean velocity, in the direction of the
mean flow. 

We would anticipate that some of the statistical properties of the true energy
dissipation defined in \eqref{epsfull} can be estimated via the 
 temporal energy dissipation defined in \eqref{epssurr}.
In doing so, we note that one needs to transition from the spatial derivatives in (\ref{epsfull}) to the 
temporal derivative in (\ref{epssurr}). This can be done under a suitable set of assumptions: Specifically, we assume that the flow is stationary, homogeneous and isotropic and that Taylor's Frozen Flow Hypothesis (TFFH), see \cite{Tay38}, holds. 
Under TFFH, we can express
spatial increments along the direction of the mean flow (in direction
$x$)  in terms of temporal increments
\begin{align*}
  v(x,y,z,t+s)-v(x,y,z,t)=v(x-\overline{v}s,y,z,t)-v(x,y,z,t).
\end{align*}
In the following, we will describe two particularly relevant ambit-based models for turbulence in more detail.

\subsubsection{Spatio-temporal models for turbulence}
Let us consider a \emph{trawl field} given by 
\begin{equation}
\Yambit =\int_{A(\mathbf{x},t) }
L(d\boldsymbol \xi,ds) = L(\Aambit),  \label{XinthdL}
\end{equation}
which is a special case of the ambit field mentioned in \eqref{eq:Cha0_ambitfield}, where the kernel function $g$ and the stochastic volatility fields $\sigma$ are assumed to be identical to one and where there is no metatime $\T$  incorporated. 
\cite{BNSch04} pointed out that, when working with an
\emph{exponentiated trawl field}, which is obtained by setting 
\begin{equation*}
X(\mathbf{x}, t) =\exp(Y(\mathbf{x}, t))=\exp(L(A(\mathbf{x}, t))),
\end{equation*}
one obtains a favourable scaling behaviour of the correlators which can be adapted to the requirements in turbulence. More precisely, we note  that the
two point correlator of order $(p,q)\in\N^2$ of $X$
is defined by
\begin{equation*}
c_{p,q}(t_{1},\mathbf{x}_{1};t_{2},\mathbf{x}_{2}) =\frac{\EE[
X(\mathbf{x}_{1}, t_{1}) ^{p}X(\mathbf{x}_{2}, t_{2}) ^{q}] }{\EE[X(\mathbf{x}_{1}, t_{1}) ^{p}] \EE[
X(\mathbf{x}_{2}, t_{2}) ^{q}] }.  
\end{equation*}
The correlator essentially extends the concept of the autocorrelation function used in time series to spatio-temporal processes.
One can show that, for exponentiated trawl fields defined above, the two point correlator is given by 
\begin{equation*}
c_{p,q}( t_{1},\mathbf{x}_{1};t_{2},\mathbf{x}_{2}) =\exp \left(c[p,q]
\leb(A(\mathbf{x}_{1}, t_{1}) \cap A(\mathbf{x}_{2}, t_{2}))\right),  
\end{equation*}
where
$c[p,q] =\mathrm{k}( p+q) -\mathrm{k}( p) -
\mathrm{k}(q)$,
and  $\mathrm{k}(\theta) :=
\log\left( \mathbb{E}\left[\e^{\theta L^{\prime }}\right]\right)$ denotes the kumulant function of the \Levy seed $L'$.
We note that $\log(c_{p,q})$ factorises into a part depending on the distribution of the \Levy seed and one depending on the size of the overlap between the ambit sets $A(\mathbf{x}_{1}, t_{1})$ and $A(\mathbf{x}_{2}, t_{2})$. This
leads to great flexibility in terms of stochastic modelling as described in 
 \cite{Schmiegel05}, \cite{SchBNEg05} and \cite{HedSch13}.
 Note that the correlators exhibit the following \emph{self-scaling property}:
\begin{equation*}
c_{p,q}( t_{1},\mathbf{x}_{1};t_{2},\mathbf{x}_{2})=
c_{1,1}(t_{1},\mathbf{x}_{1};t_{2},\mathbf{x}_{2}) ^{\bar{c}[p,q] }\,,
\end{equation*}
where $\bar{c}[p,q] =c[p,q]/c[1,1]$; such a self-scaling behaviour is widely observed in empirical studies of 
homogeneous turbulent flows.

In the stationary setting, the results can be further simplified:
Let us assume that the ambit set satisfies $\Aambit =A+(\mathbf{x},t)$, i.e.~we have a fixed set $A=A(\mathbf{0}, 0)$, at  location ${\bf 0}$  and time 0, which is then translated in space and time and whose shape does not change as it moves around. 
In that case, the trawl field defined in \eqref{XinthdL} becomes stationary and its joint cumulant function takes the form
\begin{align}\begin{split}\label{eq:JointCum}
C(\phi ,\psi; Y(\mathbf{0}, 0),\Yambit) &=
\log(\exp(i\phi Y(\mathbf{0}, 0)+ i\psi \Yambit))
\\
&=
\leb(A)\{ (c(\phi) +c(\psi)) \bar{r}(t,\mathbf{x}) )+c(\phi+\psi) r(t,\mathbf{x})\}.
\end{split}
\end{align}
Here, $c(\zeta) =C(\zeta; L^{\prime })=\log(\exp(i\zeta L'))$ denotes the cumulant function of the \Levy seed $L'$ and  
\begin{equation*}
r(t,\mathbf{x}) =\frac{\leb(A\cap \Aambit)}{
\leb(A)}, 
\end{equation*}
denotes the \emph{autodependent function} of $Y({\bf x},t)$ 
and  $\bar{r}=1-r$.
An immediate implication of \eqref{eq:JointCum} is that 
\begin{equation*}
C(\zeta; \Yambit-Y(\mathbf{0}, 0))=\leb(A)(c(-\zeta)+c(\zeta))\bar{r}(t,\mathbf{x}) .
\end{equation*}
Hence, if $L^{\prime }$ has a  symmetric law, then 
$C(\zeta;\Yambit-Y(\mathbf{0}, 0))=2\leb(A)c(\zeta) \bar{r}(t,\mathbf{x})$. 

\begin{example}
    For instance, if one works under the assumption that $L'$ has a 
 symmetric normal inverse Gaussian (NIG) distribution, then the resulting trawl field $Y(\mathbf{x},t) $ and all increments of  $Y$ also follow an NIG distribution. 
\end{example}
In the stationary case, 
we consider the two-point correlator 
\begin{equation*}
c_{p,q}(t,\mathbf{x}):=c_{p,q}( (0,\mathbf{0}) ;(t,\mathbf{x}))\,,
\end{equation*}
and note that 
$\log(c_{p,q}( t,\mathbf{x})) =\leb(A)\left(\mathrm{k}(p+q)-\mathrm{k}(p)
-\mathrm{k}(q)\right) r(t,\mathbf{x})$.
Hence, 
the self-scaling exponent $\bar{c}(p,q)$ satisfies
\begin{equation*}
\bar{c}(p,q) =\frac{\mathrm{k}(p+q)-\mathrm{k}(p)-\mathrm{k}(q)}{\mathrm{k}(2)
-2\mathrm{k}(1)}.  
\end{equation*}

We might wonder how the ambit set $A$ should be chosen to match a particular self-scaling property observed in empirical data.
 \cite{SchBNEg05} and \cite{HedSch13} tackle this question and establish that, in the stationary setting, we have 
\begin{equation*}
\frac{\partial}{\partial t}\log( c_{p,q}(t,\mathbf{x}))=\left(\mathrm{k}
(p+q)-\mathrm{k}(p)-\mathrm{k}(q)\right) 
\frac{\partial}{\partial t} \leb(A\cap(A+(\mathbf{x},t)))\,.  
\end{equation*}

Suppose for a moment that 
 $\mathbf{x}\equiv x\in\R$, and
\begin{equation}
A=\{ (\xi,s)\in\R^2 :0<s<M,\vert \xi \vert \leq
w(s) \}-(0,M)\,,  \label{Aw}
\end{equation}
where
$w(t)$ is a
nonnegative decreasing function on the interval $(0,M)$ and 
$M$ denotes  a finite decorrelation time. We get that, for any $t\in(0,M)$,
\begin{equation*}
\leb(A\cap(A+(0,t)))
=2\int_{t}^{M}w(s) \,ds.
\end{equation*}
Differentiating both sides leads to 
\begin{equation}
w(t) =-\frac{1}{2}( \mathrm{k}(2) -2\mathrm{k}
(1)) ^{-1}\frac{d}{dt}\log(c_{1,1}(t,0)).  \label{Ag}
\end{equation}
We note that the 
 derivative of $\log(c_{1,1}(t,0))$ can be estimated
from data on the timewise evolution of $Y(x,t)$ for given $x$. 
\cite{HedSch13} considered a particular example, where, for a decorrelation time $M=1$, the
function $w$ in the 
ambit set $A$  defined as in (\ref{Aw})  is given by 
\begin{equation*}
w(t)=\left(\frac{1-t^{\theta }}{1+(t/b)^{\theta }}\right)^{1/\theta }
\end{equation*}
for a constant $b\in ( 0,1)$.
Note that an  arbitrary decorrelation time $M$
can be obtained by using  the function $w(t/M)$ instead. Here,  $\theta$ denotes  a tuning parameter and $b$ represents the time scale below which the scaling behaviour has terminated. \cite{HedSch13} point out that the function $w$ is in fact its own inverse, which is well related to TFFH.  

We note that \cite{SV2024} tackled a closely related estimation problem to the one in \eqref{Ag} from a nonparametric perspective.  
Such estimation methods enable the determination of shapes of ambit sets that  are consistent with data and guarantee suitable scaling properties.

\subsubsection{Temporal models for turbulence}
A lot of research in turbulence also concentrates on the purely temporal case, where new mathematical methodology can often be developed more easily before it is further adapted to spatio-temporal settings. 
Hence, let us focus on a relevant temporal model for the energy dissipation next. 

Let $v$ denote the temporal recording of the main component of the turbulent velocity vector at a fixed spatial position. Under the TFFH, the integrated energy dissipation is
proportional to
\begin{equation*}
\int_{0}^{t}\left( \frac{dv(s) }{ds}
\right) ^{2}\,ds.
\end{equation*}
We  change the notation from $v$ to $Y$ to indicate that we are working with a stochastic process. 

However, we note that for many relevant models, such as when we assume that  $Y$ is a BSS process with  gamma kernel, see \eqref{eq:Gammakernel}, 
where $\alpha \in (-1/2,0)$, then 
$Y$ is not differentiable. 
In such a case, one could try to define the integrated energy dissipation as a suitable limit of the realised variance. I.e.~suppose we observe the process $Y$ at times $i\Delta_n$, for $i=0,\ldots,\lfloor T\Delta_n\rfloor \Delta_n$ and recall from \eqref{eq:RV} that we define the realised variance of $Y$ as  
\begin{equation*}
\RV^n(t)
=\sum_{i=1}^{\lfloor t/\Delta_n \rfloor }(\Delta_i^n Y) ^{2}, \quad \mathrm{for}\; t \in [0, T].
\end{equation*}
As discussed in Section \ref{sec:stochvolest}, in the case where $Y$ is a semimartingale,  the realised variance $\RV^n(t)$ converges in probability to the quadratic variation $[Y]_t$ of $Y_t$. Hence,  $[Y]_t$ is a natural candidate to use as the 
definition of the integrated energy dissipation.

Outside the semimartingale framework, the situation is more challenging. 
When taking the BSS process with a gamma kernel  as our underlying model, then the non-semimartingale case is obtained for
$\alpha \in ( -\frac{1}{2},0) \cup (0,\frac{1}{2}) $,
see \cite{BNSch09}.
Early work by 
 \cite{Ons49} established that Kolmogorov's statistical theory of turbulence implies that  the sample paths must
be H\"{o}lder continuous of order $\frac{1}{3}$.
It turns out that, when 
$\alpha \in (-\frac{1}{2},0)$, the sample paths of the leading term of $X$ are H\"{o}lder continuous of order 
$\alpha+\frac{1}{2}$. In particular, when 
$\alpha=-\frac{1}{6}$,
the order of H\"older continuity equals $\frac{1}{3}$. 
This finding makes this particular  BSS process an interesting contender for  a stochastic model for turbulence. It remains to address the challenge of defining the integrated energy dissipation in an appropriate way. Given the limit theory discussed in Section \ref{sec:stochvolest} and in Theorem \ref{thm:main}, in particular,  a promising approach was found to be taking the limit of an appropriately scaled realised variance as the integrated energy dissipation.

We note that much of the original limit theory developed for realised measures of BSS and related processes outside the semimartingale framework was driven by the search for a suitable definition of the integrated energy dissipation in turbulence studies. However, as the resulting limit theory introduced novel mathematical concepts and insights of independent interest, this research area has subsequently grown substantially as the references in Section \ref{sec:stochvolest} document.

\subsubsection{Further reading}
Purely spatial models based on ambit fields have been discussed in \cite{HedevangSchmiegel2014}. Also, \cite{S2020} investigates divergence and vorticity theorems for vector ambit fields and proposes a class of spatio-temporal $\R^2$-valued ambit fields with stationary and isotropic increments that  are relevant for turbulence studies. 
\cite{Bir13b} relates the Kolmogorov-Obukhov theory of turbulence to stochastic Navier-Stokes equations.
Also, we note that Ole, together with Mikkel Bennedsen and  J\"urgen Schmiegel, was working on tailoring ambit fields to turbulent boundary layer problems. 
More precisely, they were investigating
a universal statistical description of the main velocity component in a turbulent boundary layer, see \cite{turb1}; here they were exploring relevant distributions, with a particular focus on the NIG distribution, and could demonstrate that there is some ‘universality’ property in the statistical description of the turbulence.
In another project, see \cite{turb2}, they study a spatio-temporal BSS model of the main velocity component in a turbulent boundary layer. They could show that the two-dimensional correlations in the data could be described well using relatively simple shapes for the ambit sets.

\subsection{Stochastic growth models}
In spatial statistics and stochastic geometry one is interested in modelling the evolution of objects. Given a star-shaped object $S(t)\subset\mathbb R^n$ that evolves with time $t\geq 0$, one can define its boundary by the radial function
\begin{equation*}
    R(\boldsymbol{\phi},t)=\max\left\{r\,\vert\,\mathbf{z}+r(e_1(\phi),\ldots,e_n(\phi))\in S(t)\right\}.
\end{equation*}
Here, $\boldsymbol{\phi}=(\phi_1,\ldots,\phi_{n-2},\phi_{n-1})\in[0,\pi]^{n-2}\times[0,2\pi]$, $\mathbf{z}\in\mathbb R^n$ is the reference point of $S(t)$ and the functions $e_i, i=1,\ldots,n$ are defined as
\begin{align*}
    e_1(\boldsymbol{\phi})&=\cos(\phi_1), \\
    e_2(\boldsymbol{\phi})&=\sin(\phi_1)\cos(\phi_2), \\
    e_3(\boldsymbol{\phi})&=\sin(\phi_1)\sin(\phi_2)\cos(\phi_3), \\
        \vdots
     \\
    e_{n-1}(\boldsymbol{\phi})&=\sin(\phi_1)\cdots\sin(\phi_{n-2})\cos(\phi_{n-1}), \\
    e_{n}(\boldsymbol{\phi})&=\sin(\phi_1)\cdots\sin(\phi_{n-2})\sin(\phi_{n-1}). 
\end{align*}
\cite{JSVedel} suggest a growth model for star-shaped objects by modelling the speed of the 
radial function by an ambit field. In full generality, let $L(d\boldsymbol{\xi},ds)$ be a L\'evy basis on $[0,\pi]^{n-2}\times[0,2\pi]\times \mathbb R$ (hence, $d=n-1$) and $g$ a kernel function, and define the dynamical model for the radial function as
\begin{equation}
\label{radial-growth}
    \frac{\partial}{\partial t}R(\boldsymbol{\phi},t)=\mu(\boldsymbol{\phi},t)+\int_{A(\boldsymbol{\phi},t)}g(\boldsymbol{\phi},t;\boldsymbol{\xi},s)\sigma_s(\boldsymbol{\xi})L(d\boldsymbol{\xi},ds).
\end{equation}
The general trend in the growth of the star-shaped object is modelled by the deterministic function $\mu(\cdot,t):[0,\pi]^{n-2}\times[0,2\pi]\rightarrow\mathbb R$. The ambit set $A(\boldsymbol{\phi},t)\subset [0,\pi]^{n-2}\times[0,2\pi]\times(-\infty,t]$ relates past events to the current growth rate. 

The growth model in \eqref{radial-growth} makes sense also with negative values of the right-hand side, describing that the star-shaped object is shrinking. However, to have {\it growth}, one would like $\partial R(\boldsymbol{\phi},t)/\partial t$ to be non-negative. This can be achieved by considering L\'evy bases $L$ which are non-negative along with non-negative kernel functions $g$ and intermittency processes $\sigma$. Moreover, the general trend $\mu$ is also natural to assume non-negative in this case.

\subsubsection{A tumour growth model in the plane}
In \cite{JSVedel}, a growth model on the plane $\mathbb R^2$ (i.e., $n=2$) is studied in detail based on a special case of the ambit field model in \eqref{radial-growth}. \cite{JSVedel} suppose that the L\'evy basis is only depending on the angle, $L(d\mathbf{\xi})$ and the kernel function is therefore not depending on $s$. Furthermore, 
intermittency is set to $\sigma=1$. Since $n=2$, we have $\phi\in[0,2\pi]$, with $e_1(\phi)=\cos(\phi)$ and $e_2(\phi)=\sin(\phi)$. Thus, their planar tumour growth model is 
\begin{equation}
\label{planar-radial}
    \frac{\partial}{\partial t}R(\phi,t)=\mu(\phi,t)+\int_{A(\phi,t)}g(\phi,t;\mathbf{\xi})L(d\mathbf{\xi}).
\end{equation}
We now take a closer look at some of their modelling choices and analysis for the model \eqref{planar-radial}.

The random field describing the evolution of the 
radial function $R$ is given by
\begin{align}
\label{radial-explicit}
R(\phi,t)&=R_0(\phi)+\int_0^t\mu(\phi,s)ds+\int_0^t\int_{A(\phi,s)}g(\phi,s;\xi)L(d\xi)ds \nonumber \\
    &=R_0(\phi)+\overline{\mu}(\phi,t)+\int_{\overline{A}(\phi,t)}\overline{g}(\phi,t;\xi)L(d\xi).
\end{align}
Here, $\overline{\mu}(\phi,t)=\int_0^t\mu(\phi,s)ds$,
\begin{equation*}
  \overline{A}(\phi,t)=\bigcup_{0\leq s\leq t}A(\phi,s),  
\end{equation*}
and
\begin{equation*}
    \overline{g}(\phi,t;\xi)=\int_0^t\mathbf{1}_{A(\phi,s)}g(\phi,s;\xi)ds.
\end{equation*}
To commute the integrals to reach \eqref{radial-explicit} we need to appeal to a stochastic Fubini theorem. We refer to \citet[Prop. 45]{BNBV2018_ambitbook} for a Fubini theorem for ambit fields and for precise conditions that allows for commuting the $ds$-integral with that of $L(d\xi)$.

The specification of the ambit set is an important piece in the growth model specification. \cite{JSVedel} suggest for example a model for the sphere of influence where a minimal time lag is introduced in the time-component,
\begin{equation*}
A(\phi,t)\subset[0,2\pi]\times[t-\tau(t),t],
\end{equation*}
where $t\mapsto \tau(t)\geq 0$ is some measurable function. One may let this function be constant, $\tau(t)=\delta$, meaning that the sphere of influence goes back a fixed time step $\delta$ with respect to current time $t$. \cite{JSVedel} also suggest to specify $t\mapsto t-\tau(t)$ being non-decreasing, which, in other words, means to have $\tau$ such that $t_1-t_2\geq \tau(t_1)-\tau(t_2)$ for any $t_1\geq t_2$. For such ambit sets, \citet[Prop. 7 and Ex. 8]{JSVedel} compute explicit 
expressions for the covariance between 
$R(\phi_1,t_1)$ and $R(\phi_2,t_2)$ given the Fourier representation of the kernel function $g$. The covariance is in general non-separable. However, it is separable in special cases of the kernel function $\overline{g}$, i.e., when the Fourier expansion of the kernel function only consists of its first basis term, in which case
\begin{equation*}
    \text{Cov}(R(\phi_1,t_1), R(\phi_2,t_2))\sim\cos(\phi_1-\phi_2)h(t_1,t_2),
\end{equation*}
for some function $h$ depending on $\tau$ and the kernel function. If $\tau(t)=\delta$ is constant, then $h(t_1,t_2)=h(t_1-t_2)$ and we obtain a covariance function being stationary in angle and time. We refer to \cite[Prop.~7]{JSVedel} for details, where also a possible restriction in the angular dimension of the
ambit set is considered (see Sect. 5.2 in that paper). Noteworthy is the close link between the dependency structures obtained by a carefully selected ambit set and kernel function, and typical covariance functions appearing for other modeling approaches in stochastic geometry (for instance, see Ex. 10 and the discussion following in \cite{JSVedel}).

\subsubsection{Further reading}
Definitions of L\'evy-based growth models based on ambit fields appeared already in \cite{Schmiegel06} and 
\cite{BNSch07a}. \cite{JV} studied a discrete-time version of the growth model in \eqref{planar-radial} with a Gaussian L\'evy basis. Interestingly, ambit fields taking 
non-negative values are used in intermittency and 
volatility modelling, see more about this in \cite{BNBV2018_ambitbook}.

\section{Conclusion and outlook}\label{sec:con}

Ambit fields constitute a rich and analytically tractable class of random fields that are suitable for modelling spatio-temporal phenomena. 
We recall that four aspects of ambit stochastics set it apart from much of
the traditional literature on stochastic analysis: 
First, the inclusion of stochastic volatility/intermittency
as an integrand allows for stochastic scaling of the noise component, providing great flexibility in modelling. 
Second, the concept of metatimes naturally extends the idea of a time change of a stochastic process to spatio-temporal settings and constitutes a complementary method for incorporating stochastic volatility.
Third, 
 ambit sets are a new concept used to describe the region of interest where random shocks need to be taken into account.
 Fourth, ambit processes and fields can allow for both semimartingale and non-semimartingale settings and can therefore be tailored to a wide range of applications.

 Ole's work on ambit stochastics had a lasting impact on probabilistic and statistical theory, methodology, and a wide range of applications,  with turbulence always being of particular interest to Ole. 

Research in ambit stochastics has developed further recently, so let us briefly outline some of the current trends and directions. This outlook is not intended to be exhaustive.


{\bf Inference for multivariate and  high-dimensional settings:} The asymptotic theory presented in Section \ref{sec:stochvolest} predominantly focused on one-dimensional settings. In recent years, the theory was further adapted to multivariate settings, see e.g.~\cite{HP2021}.
Also, going back to the  Ornstein-Uhlenbeck process, which is a classical example within the class of \Levy semistationary processes, the question of drift estimation in high-dimensional settings was tackled, see e.g.~\cite{CMP2020},
and its extensions to general diffusion settings, see  \cite{CMP2024}.


Related work includes a network setting or graph structure into Ornstein-Uhlenbeck (and related) processes. 
In this realm, \cite{CV22, CV22b} introduced the graphOU process, which was extended to multivariate and graph continuous time autoregressive processes in \cite{LPV2023}. Also, very much in the spirit of Ole's and Robert Stelzer's work on multivariate supOU processes, see \cite{BNStelzer2011}, a graph version of such processes has been proposed in  \cite{MV2024}. 
Also, cluster algorithms for determining groups in graphOU processes are currently being developed in \cite{NV2024}.

Beyond the high-dimensional setting, work on ambit fields has been extended to the infinite-dimensional setting, see \cite{BE2015}. Also,  \cite{BS2018} developed continuous-time autoregressive moving-average  processes in Hilbert spaces, which are connected to higher-order SPDEs like the stochastic wave equation. 

{\bf Software:} 
For researchers  interested in using ambit fields in their own work, we mention the recent {\tt R}
 packages {\tt trawl, ambit}\footnote{ \cite{ambit,trawl}}, which are available on CRAN,  and the Python library {\tt Ambit-Stochastics}\footnote{\cite{Leonte24}}, which contains code for simulation and inference of ambit fields.

\vspace{1cm}
{\bf Acknowledgement:} Parts of this article were presented as a plenary talk at the Ole E.~Barndorff-Nielsen Memorial Conference  held at Aarhus University from 29 to 31 May 2024. We would like to thank the local organisers for hosting us.
Also, we thank Mikkel Bennedsen and Mark Podolskij for their detailed feedback and suggestions on an earlier draft of this article.   
AEDV has been supported   by EPSRC through the grant \emph{Network Stochastic Processes and Time Series (NeST)}, EP/X002195/1.
\bibliographystyle{agsm}
\bibliography{AmbitSurvey}
\end{document}